\documentclass[11pt,letter]{article}
\usepackage{deauthor}
\usepackage{times}
\usepackage{graphicx}
\usepackage{amsfonts}   
\usepackage{amsmath}
\usepackage{amssymb}
\usepackage{booktabs}
\usepackage{multirow}
\usepackage[font=small,labelfont=bf]{caption}

\usepackage{deauthor}

\usepackage[utf8]{inputenc} 
\usepackage[T1]{fontenc}    
\usepackage{hyperref}       
\usepackage{url}            
\usepackage{booktabs}       
\usepackage{amsfonts}       
\usepackage{nicefrac}       
\usepackage{microtype}      
\usepackage{lipsum}

\usepackage{graphicx}
\usepackage{algorithm}
\usepackage{caption}
\usepackage{algorithmicx}
\usepackage{algcompatible}
\usepackage{multirow}

\usepackage{hyperref}

\begin{document}

\title{Predicting Diffusion Reach Probabilities via Representation Learning on Social Networks}

\def\sharedaffiliation{%
\end{tabular}
\begin{tabular}{c}}
\author{
Furkan Gursoy\\
\texttt{furkan.gursoy@boun.edu.tr}
\and
Ahmet Onur Durahim \\
\texttt{onur.durahim@boun.edu.tr}
\vspace{10pt}
\sharedaffiliation
Dept. of Management Information Systems\\
  Boğaziçi University\\
  Istanbul, Turkey
}

\maketitle

\begin{abstract}
Diffusion reach probability between two nodes on a network is defined as the probability of a cascade originating from one node reaching to another node. An infinite number of cascades would enable calculation of true diffusion reach probabilities between any two nodes. However, there exists only a finite number of cascades and one usually has access only to a small portion of all available cascades. In this work, we addressed the problem of estimating diffusion reach probabilities given only a limited number of cascades and partial information about underlying network structure. Our proposed strategy employs node representation learning to generate and feed node embeddings into machine learning algorithms to create models that predict diffusion reach probabilities. We provide experimental analysis using synthetically generated cascades on two real-world social networks. Results show that proposed method is superior to using values calculated from available cascades when the portion of cascades is small.
\end{abstract}

\keywords{Social Networks \and Information Networks \and Representation Learning \and Influence Maximization}

\section{Introduction}
Social networks play an increasingly important role in our daily lives, and in the world in general. Politics, business, and other matters of our lives are significantly shaped by social influence that we are exposed to via the social networks we are part of. The world has evolved to be a place where information can spread very quickly and easily through cascades. Information cascades in social networks can be modeled by stochastic models such as Markov random fields (Domingos \& Richardson, 2001), Independent Cascade Model (ICM), Linear Threshold Model (LTM) (Kempe, Kleinberg, \& Tardos, 2003), or deterministic models such as Deterministic LTM (Gursoy \& Gunnec, 2018). In this work, we employ ICM which is one of the most commonly studied diffusion model in the literature.

Independent Cascade Model (ICM) assumes that diffusion time steps are discrete. At any time, a node can be either active (i.e., influenced) or inactive. An active node may attempt to activate a neighboring inactive node only once, and a node cannot become inactive later once it is active (i.e., a progressive model). The process starts with some initially active nodes which serve as the seed nodes. A node $v$ that is activated at time $t$ tries to activate its inactive neighbor nodes at time $t+1$. The attempt is successful with probability $p_{vu}$. The process runs until the time step where no more nodes get activated.

\begin{figure}[]
\includegraphics[width=9cm]{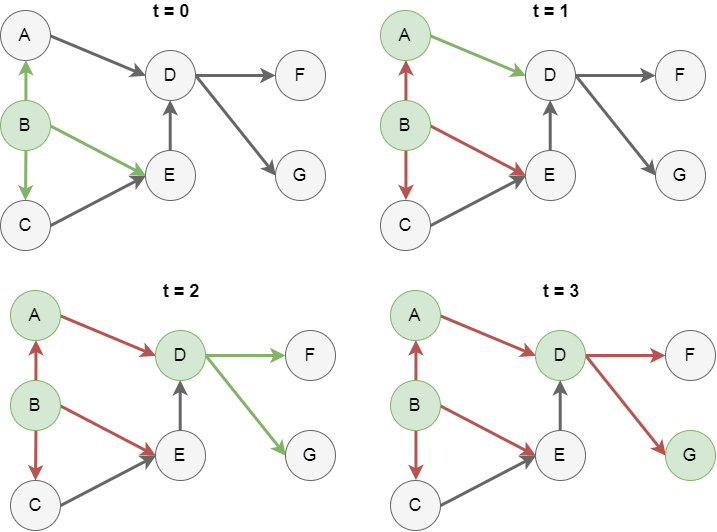}
\centering
\caption{Independent Cascade Model}
\label{fig:icm}
\end{figure}

Figure \ref{fig:icm} illustrates the diffusion process in ICM. $B$ is selected as the seed node and activated at $t = 0$. It then attempts to activate its neighbors. $A$, $C$, and $E$ have activation probabilities of $p_{BA}$, $p_{BC}$, and $p_{BE}$ respectively. At $t = 1$, only $A$ is activated by $B$; and $B$ cannot activate any of its neighbors anymore. $A$ then proceeds to activate $D$ in a similar fashion. After $D$ attempts to activate its neighbors and activate $G$, the diffusion terminates since there does not remain any active node which can attempt to activate neighbors.

Diffusion reach probability between two nodes can be defined as the probability of the cascade originating from a node reaching to the other node, by traversing the nodes between them if there are any. An infinite number of cascades would enable calculation of diffusion reach probabilities, however there exists only a finite number of cascades, and in general only a limited portion of them are accessible. Predicting such probabilities is, therefore, a nontrivial problem. 

Machine learning might be utilized in such prediction tasks. However, most machine learning algorithms require input data to be in a tabular form in which rows represent cases and columns represent feature space. The choice of the feature set has a significant effect on the performance of a machine learning model. Hence, a considerable amount of effort is actually spent on engineering better features. Representation learning is a way of automatically discovering important features which replaces the time consuming manual feature engineering. Representation learning in social networks is concerned about finding learning methods which can embed nodes to a latent space in a way that the resulting embeddings contain maximum information within a reasonable dimensionality. These learned latent features then can be used in machine learning tasks.

\section{Related Work}\label{sec:walk}
Apart from the earlier matrix factorization methods, seminal algorithm of Perozzi, Al-Rfou, \& Skiena (2014) named \textit{DeepWalk} paved the way for most of the future studies. \textit{DeepWalk} is a random walk-based method, to learn latent representations of vertices in a network by optimizing the probability of nodes occurring in the same random walk using gradient descent. Tang et al. (2015) proposed Large-scale Information Network Embedding (LINE) which considers first-order proximity (i.e., sharing a tie) in addition to second-order proximity (i.e., sharing the neighborhood). This way, \textit{LINE} improved \textit{DeepWalk} which only considers second-order proximity. Improving this line of work even further, Grover \& Leskovec (2016) proposed \textit{node2vec} which generates biased random walks by trading-off between depth-first search and breadth-first search. Hence, \textit{node2vec} is able to discover a more diverse neighborhood structure. 

There are number of other studies which considers context of links and nodes (Tu, Liu, Liu, \& Sun, 2017), node attributes (Huang, Li, \& Hu, 2017); or focuses on developing a meta-strategy (Chen, Perozzi, Hu, \& Skiena, 2017). On the other hand, Tu, Zhang, Liu \& Sun (2016) and Wang, Cui, \& Zhu (2016) develop semi-supervised representation learning algorithms.

Another line of work in representation learning is concerned specifically with its applications related to information diffusion in social networks. Works of Bourigault, Lamprier, \& Gallinari (2016) and Li, Ma, Guo, \& Mei (2017) are among such studies which aims to predict diffusion probabilities or make use of existing information diffusion cascades in creating latent representations.

Overall, the use of representation learning for information diffusion in social networks is a promising field. There is an ample space for development of more sophisticated algorithms, extending the existing problems to cover different problem settings, and applications of those methods in different fields including interdisciplinary works.

\section{Methodology}

The main objective of this study is to predict the probability of a new diffusion cascade, on a given a directed social network $G(V,E)$, originating from a node $u$ reaching to any other node $v$ given partial information on underlying network structure, only the connections and directions but not the strengths, and limited number of cascades on the actual network. Here, the $i$th cascade, $C_i$, contains timestep-node pairs $(t, V_t)$ where $V_t$ contains nodes $v \in V_t$ that are activated at time $t$.

\subsection{Dataset Generation}

Diffusion cascades are synthetically created under ICM using real-world graph datasets. Email-Eu-core (Leskovec, Kleinberg, \& Faloutsos, 2007) is a directed network generated from internal email data of a large European research institution. There is an edge $(u, v)$ in the network if person $u$ sent person $v$ an email. Bitcoin Alpha (Kumar, Spezzano, Subrahmanian, \& Faloutsos, 2016) is the who-trusts-whom directed network of people who trade using Bitcoin on a platform called Bitcoin Alpha. Number of nodes and links for each network is shown in Table \ref{tab:graphs}.

\begin{table}[b]
\centering
\caption{Graphs used in the study}
\small
\small
\setlength\tabcolsep{2.5pt}
\label{tab:graphs}
\begin{tabular}{|c|c|c|}
\hline
                       & \textbf{\# of nodes} & \textbf{\# of links} \\ \hline
\textbf{Email-Eu-core} & 1005                 & 25,571                \\ \hline
\textbf{Bitcoin Alpha} & 3783                 & 24,186               \\ \hline
\end{tabular}
\end{table}

Diffusion cascades are created in the following way. We assign activation probabilities $p_{uv}$, the probability that node $u$ activates node $v$, randomly between 0 and a selected value $max_p$. Then, for each node $v \in V$, number $r$ of cascades are initiated by activating $v$. The diffusion takes place according to the ICM. In this way, total of $|V|r$ cascades are generated for each dataset.

Note that, the activation probabilities which are assigned for generating diffusion cascades are assumed to be unknown in the rest of the study. As far as the rest of the study concerns, the cascades are treated to be generated by some unknown processes or taken from real-world cascades.

\subsection{Proposed Method}

Our proposed method aims to predict diffusion probabilities between any two nodes regardless of whether they are neighbors or not. Thus, to measure accuracy, we need actual and predicted values. The actual diffusion probability between $u$ and $v$ is assigned as $P(v | V_{*,0} = \{u\})$ that is the probability of $v$ occurring in a cascade started from $u$. The algorithm for calculating actual reach probabilities is given in Algorithm \ref{alg1}. Note that, as $r$ approaches to infinity, our actual values approach to their true values. Therefore, it is important to make the distinction between true values, actual values, label values, and predicted values. Actual values are indeed estimations of true values. Since we assume that we can not precisely know the true activation values, computing the actual values from all generated cascades allows us to obtain some accuracy scores by comparing them with the predicted values. Label values, on the other hand, are calculated using only a portion of all cascades and used in training of the model. Label values are calculated in the same way as actual values.

\begin{algorithm}[t]
\caption{Actual Diffusion Probability Calculation}
\label{alg1}
\begin{algorithmic}[1]

\REQUIRE {$C$: a cascade list  where $C[i][j][k]$ is the $k^{th}$ node activated at time $j$ for the cascade $i$, $x$: number of cascades,  $y$: number of timesteps a particular cascade diffuses for, $z$: number of nodes a cascade contains at a particular timestep, $r$: number of cascades for each node} 
\STATE $M \gets$ a zero matrix to count how many times each node appeared in all cascades for given seed nodes 
\FOR { $i = 1$ to $x$ }
    \FOR { $j = 1$ to $y$ }
        \FOR { $k = 1$ to $z$ }
            \STATE $M[C[i][1][1]],[C[i][j][k]] += 1$
        \ENDFOR
    \ENDFOR
\ENDFOR

\FOR { $i = 1$ to $|V|$ }
    \FOR { $j = 1$ to $|V|$ }
       \STATE $A[i,j] = M[i,j]/r$
    \ENDFOR
\ENDFOR
\ENSURE $A$, where $A[i,j]$ is the actual diffusion probability from node $i$ to node $j$
\end{algorithmic} 
\end{algorithm}

The node embeddings are generated by \textit{node2vec} algorithm. Given a node embedding $e_v$ for node $v$ and node embedding $e_u$ for node $u$, link embedding $e_{vu}$ is created by concatenating $e_v$ and $e_u$. Note that, link embeddings are created for every node pair regardless of whether they are neighbors in the network or not.

Machine learning models from Python's \href{http://scikit-learn.org}{\textit{scikit-learn}}
 library are then used to train the models. Link embeddings are used as features and corresponding label probabilities are used as target values. There are as much as $|V|(|V|-1)$ link embeddings since there is an embedding for each node pair.
 
 The source code and datasets employed in this study are available at \href{https://furkangursoy.github.io}{\textit{furkangursoy.github.io}}

\section{Experimental Results}

For each graph, two cascade sets are generated: one with random activation probabilities between $0$ and $0.05$, and the other one with probabilities between $0$ and $0.1$. Therefore, we obtain a total of 4 different cascade sets. Each cascade set contains $|V|r$ different cascades where $r=20$ for the purposes of our experimental analysis.

For each cascade set, label values are calculated based on portions of all cascades in the given cascade set. $10\%$, $20\%$, $40\%$, and $60\%$ of all cascades are considered. The cascades are selected randomly from all cascades, resulting in different number of cascades for different seed nodes. This effectively results in 4 different sets of label values (e.g., diffusion probabilities) for each cascade set. The intuition behind this is that; one has access only to a portion of all information cascades in real life. In total, we obtain 16 datasets to train our machine learning models.

Parameters of \textit{node2vec} are set as follows: directed network, 128 dimensions (i.e., number of latent features), walk length of 20, and window size of 5. 

Performance results are measured by Mean Absolute Error (MAE) and presented in Table \ref{tab:perfresults}. In the table, BM is an abbreviation for Benchmark and its value is calculated by comparing label values with actual probabilities. GrdBst is an abbreviation for Gradient Boosting, MLP is an abbreviation for Multilayer Perceptron. Performances of GrdBst and MLP is calculated by comparing predicted values with actual probabilities. Since it takes excessively long time for GrdBst to train, GrdBst experiments are not performed for Bitcoin Alpha.

\begin{table*}[h!]
\centering
\small
\small
\setlength\tabcolsep{2.5pt}
\caption{Performance Results (MAE scores)}
\label{tab:perfresults}
\begin{tabular}{cc|r|r|r|r|r|}
\cline{3-7}
                                                 &                 & \multicolumn{3}{c|}{Email-Eu-core}                                               & \multicolumn{2}{c|}{Bitcoin Alpha}                 \\ \hline
\multicolumn{1}{|c|}{Activation Prob.}           & Cascade Portion & \multicolumn{1}{c|}{BM} & \multicolumn{1}{c|}{GrdBst} & \multicolumn{1}{c|}{MLP} & \multicolumn{1}{c|}{BM} & \multicolumn{1}{c|}{MLP} \\ \hline
\multicolumn{1}{|c|}{\multirow{4}{*}{(0, 0.05)}} & 10\%            & 0.0874                  & \textbf{0.0538}             & 0.0592                   & \textbf{0.0013}         & 0.0017                   \\ \cline{2-7} 
\multicolumn{1}{|c|}{}                           & 20\%            & 0.0667                  & \textbf{0.0508}             & 0.0514                   & \textbf{0.0012}       & 0.0015               \\ \cline{2-7} 
\multicolumn{1}{|c|}{}                           & 40\%            & \textbf{0.0419}         & 0.0498                      & 0.0478                   & \textbf{0.0009}         & 0.0012                   \\ \cline{2-7} 
\multicolumn{1}{|c|}{}                           & 60\%            & \textbf{0.0272}         & 0.0492                      & 0.0460                   & \textbf{0.0006}         & 0.0014                   \\ \hline
\multicolumn{1}{|c|}{\multirow{4}{*}{(0, 0.1)}}  & 10\%            & 0.1860                  & \textbf{0.1118}             & 0.1445                   & 0.0210                  & \textbf{0.0176}          \\ \cline{2-7} 
\multicolumn{1}{|c|}{}                           & 20\%            & 0.1219                  & \textbf{0.0992}             & 0.1055                   & 0.0177                  & \textbf{0.0164}          \\ \cline{2-7} 
\multicolumn{1}{|c|}{}                           & 40\%            & \textbf{0.0707}                 & 0.0959                      & 0.0850                   & \textbf{0.0118}         & 0.0163                   \\ \cline{2-7} 
\multicolumn{1}{|c|}{}                           & 60\%            & \textbf{0.0459}                  & 0.0952                      & 0.0867                   & \textbf{0.0076}         & 0.0175                   \\ \hline
\end{tabular}
\end{table*}

If the portion rate was 100\%, Benchmark error would be zero by definition. Accordingly, as the ratio of available cascades increases, performance of Benchmark gets better. However, when the available portion of cascades is smaller, machine learning algorithms perform better than the benchmark; which confirms our hypothesis that information extracted from the underlying network structure helps in predicting the diffusion reach probabilities in case of having only a portion of all cascades.

In experiments for Email-Eu-core dataset, when 10\% of cascades are available, error rate of GrdBst is 38\% to 40\% lower than error rate of BM whereas error rate of MLP is 22\% to 32\% lower. When available cascade portion is 20\%, error rates of GrdBst and MLP are 19\% to 24\% and 14\% to 23\% lower than error rate of BM, respectively. Similar results hold for experiments on Bitcoin Alpha network when activation probabilities are set between $0$ and $0.1$. In both networks, when available portion of cascades increases beyond 40\%, Benchmark begins to outperform our strategy. The experiments on Bitcoin Alpha with activation probabilities between $0$ and $0.1$ seem to be an exception among all experimental results. Therefore, as desired, our strategy works well for the cases where number of cascades is limited.

When performances of GrdBst and MLP methods are compared in Email-Eu-core experiments, it can be seen that MLP has a more varying performance compared to GrdBst when available cascade portions change. Thus, the performance of GrdBst is less dependent on the portion of cascades whereas performance of MLP is influenced more by it. MLP outperforms GrdBst when there is a larger number of cascades, however; in the cases where MLP performs better than GrdBst, BM outperforms both.

Experiments are performed on a computer with Intel Xeon CPU @ 2.40 GHz, and 64 GB memory. The average runtimes for training the learning models is given in Table \ref{tab:runtimes}.  In addition to differences between the two algorithms, runtime of MLP is shorter since it works in parallel whereas GrdBst works on a single core. On the other hand, the large runtime differences between the two graphs are due to the size of the training data. Number of rows in the training data is proportional to the square of number of nodes. Accordingly, training data from Bitcoin Alpha has approximately 14 times more rows compared to that of Email-Eu-core.

\begin{table}[h!]
\centering
\small
\small
\setlength\tabcolsep{2.5pt}
\caption{Runtime Results}
\label{tab:runtimes}
\begin{tabular}{|c|c|c|c|}
\hline
\multicolumn{2}{|c|}{Email-Eu-core}  & \multicolumn{2}{c|}{Bitcoin Alpha} \\ \hline
GrdBst          & MLP                & GrdBst          & MLP              \\ \hline
$\approx$ 2 hours & $\approx$ 10 minutes & $>$ 1 day   & $\approx$ 2.5 hours  \\ \hline
\end{tabular}
\end{table}

\section{Conclusion}\label{sec:conclusion}

In this work, we proposed a strategy which utilizes representation learning for predicting diffusion reach probabilities between nodes using the available cascade information on the network. Our work is novel in a sense that it aims to predict diffusion probabilities between any nodes regardless of whether they are neighbors or not, in comparison to previous literature where probabilities between neighboring nodes are of concern. Also, utilization of representation learning in diffusion probability prediction is, to the best of our knowledge, a novel strategy.

Experimental analyses showed that MAE of our method is up to 40\% lower than that of the benchmark when available portion of cascades is 10\%, and up to 24\% lower when portion of cascades is 20\%. Hence, our novel strategy stands out as a promising method in predicting diffusion reach probabilities when only partial information about cascades and underlying network structure is available.

Once the diffusion probabilities between all nodes are predicted, one can utilize this information to estimate diffusion cascades and design viral marketing campaigns accordingly. For instance, in deciding which seed set to hire among the candidate seed sets for an influencer marketing campaign in an online social network, the marketers can compare the estimated future cascades originated from these different seed sets. This can also be used in targeted campaigns where the final cascades are compared based on desired characteristics the nodes in those cascades have. 

Furthering the current work, we plan to improve on two aspects of this methodology. First, generation of link embeddings can be improved in a way that the latent representation contain more and better suiting information for the problem of diffusion reach probability prediction. Second, created models can be improved by experimenting with variety of machine learning algorithms as well as by designing a better training data (e.g., in terms of imbalance). Moreover, scalable methods should be developed to accommodate larger networks.

\section{Acknowledgement}\label{sec:conclusion}
This research was partially supported by Bogazici University Research Fund (BAP), Project Number: 15N03SUP2.

\newpage
\section*{References}

Bourigault, S., Lamprier, S., \& Gallinari, P. (2016). Representation learning for information diffusion through social networks.\textit{ Proceedings of the Ninth ACM International Conference on Web Search and Data Mining - WSDM 16}. \vspace{4pt}

\noindent Chen, H., Perozzi, B., Hu, Y., \& Skiena, S. (2017). HARP: hierarchical representation learning for networks. \textit{arXiv preprint arXiv:1706.07845}.\vspace{4pt}

\noindent Domingos, P., \& Richardson, M. (2001). Mining the network value of customers. \textit{Proceedings of the Seventh ACM SIGKDD International Conference on Knowledge Discovery and Data Mining - KDD 01}.\vspace{4pt}

\noindent Grover, A., \& Leskovec, J. (2016). Node2vec: Scalable feature learning for networks. \textit{Proceedings of the 22nd ACM SIGKDD International Conference on Knowledge Discovery and Data Mining - KDD 16}.\vspace{4pt}

\noindent Gursoy, F., \& Gunnec, D. (2018). Influence maximization in social networks under deterministic linear threshold model. \textit{Knowledge-Based Systems}.\vspace{4pt}

\noindent Huang, X., Li, J., \& Hu, X. (2017). Accelerated attributed network embedding. \textit{Proceedings of the 2017 SIAM International Conference on Data Mining}, 633-641. \vspace{4pt}

\noindent Kempe, D., Kleinberg, J., \& Tardos, É. (2003). Maximizing the spread of influence through a social network.\textit{ Proceedings of the Ninth ACM SIGKDD International Conference on Knowledge Discovery and Data Mining - KDD 03}. \vspace{4pt}

\noindent Kumar, S., Spezzano, F., Subrahmanian, V. S., \& Faloutsos, C. (2016). Edge weight prediction in weighted signed networks. \textit{2016 IEEE 16th International Conference on Data Mining (ICDM)}. \vspace{4pt}

\noindent Leskovec, J., Kleinberg, J., \& Faloutsos, C. (2007). Graph evolution: Densification and shrinking diameters. \textit{ACM Transactions on Knowledge Discovery from Data, 1(1)}.\vspace{4pt}

\noindent Li, C., Ma, J., Guo, X., \& Mei, Q. (2017). DeepCas: an end-to-end predictor of information cascades. \textit{Proceedings of the 26th International Conference on World Wide Web - WWW 17}.\vspace{4pt}

\noindent Perozzi, B., Al-Rfou, R., \& Skiena, S. (2014). Deepwalk: Online learning of social representation. \textit{Proceedings of the 20th ACM SIGKDD International Conference on Knowledge Discovery and Data Mining - KDD 14}.\vspace{4pt}

\noindent Tang, J., Qu, M., Wang, M., Zhang, M., Yan, J., \& Mei, Q. (2015). Line: Large-scale information network embedding. \textit{Proceedings of the 24th International Conference on World Wide Web - WWW 15}.\vspace{4pt}

\noindent Tu, C., Liu, H., Liu, Z., \& Sun, M. (2017). CANE: Context-aware network embedding for relation modeling. \textit{Proceedings of the 55th Annual Meeting of the Association for Computational Linguistics (Volume 1: Long Papers)}.\vspace{4pt}

\noindent Tu, C., Zhang, W., Liu, Z., \& Sun, M. (2016, July). Max-Margin DeepWalk: Discriminative learning of network representation. \textit{In IJCAI} (pp. 3889-3895).\vspace{4pt}

\noindent Wang, D., Cui, P., \& Zhu, W. (2016, August). Structural deep network embedding. \textit{In Proceedings of the 22nd ACM SIGKDD international conference on Knowledge discovery and data mining} (pp. 1225-1234). ACM.\vspace{4pt}
\end{document}